\documentstyle[twocolumn,prb,epsfig,aps]{revtex}
\begin{document}
\draft
\title{Thermodynamic and Transport Properties of Superconducting $Mg^{10}B_2$}
\author{ D. K. Finnemore,  J. E. Ostenson,  S. L. Bud'ko, G. Lapertot,\footnote {On 
leave from Commissariat a l'Energie Atomique, DRFMC-SPSMS, 38054 Grenoble, France} 
 P. C. Canfield  }
\address{Ames Laboratory, U.S. Department of Energy and Department of Physics and Astronomy\\
Iowa State University, Ames, Iowa 50011}

\date{\today}
\maketitle
\begin{abstract}
Transport and thermodynamic properties of a sintered pellet of 
the newly discovered $MgB_2$ superconductor have been measured to determine the 
characteristic critical magnetic fields  and critical current densities.  Both 
resistive transition and magnetization data give similar values of the upper critical field, $H_{c2}$,  with magnetization data giving $dH_{c2}/dT=0.44~T/K$ at the transition temperature of $T_c=40.2~K$.  Close to the transition temperature, magnetization 
curves are thermodynamically reversible, but at low temperatures the trapped flux can be 
on the order of one Tesla.  The value of $dH_c/dT$ at $T_c$ 
is estimated to be about $12~mT/K$, a value similar to classical superconductors like Sn.  Hence, the Ginsburg-Landau parameter $\kappa \sim 26$.  Estimates of the critical supercurrent density, $J_c$,  
using hysteresis loops and the Bean model give critical current densities on the order 
of $10^5~A/cm^2$.   Hence the supercurrent coupling through the grain boundaries is 
comparable to intermetallics like $Nb_3Sn$.  
\end{abstract}


\section{Introduction}

With the discovery of superconductivity in $MgB_2$ at about $39~K$ by Akimitsu and 
co-workers,\cite {1} there is an opportunity to study superfluid transport and phase locking of 
conduction electrons in a whole new class of materials.  In early studies of the $B$ 
isotope effect,\cite {2} Bud'ko and co-workers found the superconducting transition temperature, 
$T_c$, increased from $39.2~K$ for $Mg^{11}B_2$ to $40.2~K$ for $Mg^{10}B_2$ giving 
a partial isotope exponent $\alpha _B=0.26$ for the 
isotope relation $T_c\sim M^{-\alpha _B }$ where $M$ is the isotope mass.  This is a clear indication 
that the phonons are playing an important role in the superconducting interaction.  In addition, 
band structure calculations\cite {3} indicate a rather isotropic electrical transport in spite of 
the very layered appearance of honeycombed boron and hexagonal 
magnesium networks in the material.  

The purpose of this letter is to report measurements of the critical fields of this material 
to get an estimate of the characteristic length and energy scales for comparison with the  classical superconductors like Nb and the high temperature superconductors, like 
$YBa_2Cu_3O_{7-\delta }$.  A second goal is to obtain some measure of the supercurrent 
transport through grain boundaries for comparison with the high
 temperature superconductors where there are serious weak link problems with the grain 
boundaries.

\section{Experiment}

A sintered pellet of $Mg^{10}B_2$ was made by sealing a stoichiometric mixture of 
$Mg$ and $^{10}B$ in a Ta tube and heating to $950 ^\circ C$ for two hours.\cite {2}
Magnetization was measured with a Quantum Designs SQUID magnetometer with a $60~mm$ scan length on a $14.22~mg$ piece from the same batch of $Mg^{10}B_2$ material reported 
previously.\cite {2}  Electrical resistance between $1.9~K$ and $300~K$ in applied magnetic 
field up to $9~T$ was measured in a Quantum Designs PPMS-9 apparatus 
on another piece from the same $Mg^{10}B_2$ batch using standard four probe ac 
resistance technique at $f~=~16~Hz$ and a current density of $0.1-0.3~A/cm^2$.  Electrical 
contacts were made with Epo-tek $H20E$ silver epoxy.  

\section{Results and discussion}
The temperature dependent electrical resistance of the material from $300~K$ to $1.9~K$, shown in Fig. 1, 
roughly obeys $R~=~R_0~+~R_1T^3$ power law in the normal state.  The residual 
resistance ratio (RRR) is approximately $20$, which is relatively high for a 
sintered polycrystalline sample.   Near $T_c$, 
the data show considerable flux-flow broadening as shown by the inset in Fig. 1 where the 
scans range from $0$ to $9~T$. 
 Transition widths gradually broaden from $0.5~K$ at $0$ field to $9.5~K$ at $\mu _oH=9~T$ indicating a broad region 
of flux-flow resistivity.\cite {4}  The extent of  this flux-flow regime is delineated by the 
onset and completion temperatures shown in  Fig. 2 by the vertical dash lines.  

Possibly the most interesting aspect of these data is 
the strong magnetoresistance which shows the resistance at $45~K$ rising approximately 
$80$ percent in $9~T$.  If these data are plotted on a Kohler Plot shown in Fig. 3, the 
$\Delta \rho/\rho _o~vs.~H/\rho _o$ curve is a straight line for the change of temperature at 
constant field (open circles) and for changing field at constant temperature (solid triangles).  
This observation is consistent with the observed magnetoresistance being a band effect and 
intrinsic to the sample.

Magnetization data are exemplified by the $36~K$ and $30~K$ runs in Fig.4. For the $30~K$ 
run, the magnetization abruptly departs from the background at $H_{c2}=4.4~T$ and is 
reversible to an accuracy of one percent of the magnetization down to $H_{irr}=1.6~T$. 
Because the background magnetization is very small, linear in magnetic field, and independent of temperature between $42~K$ and $50~K$, this correction is easily made and $H_{c2}$ and 
$H_{irr}$ are fairly easy to measure.  At $7~T$, the magnitude of the background is 
about $0.13~emu/cm^3$ and the correction can be made to an accuracy of $0.01~emu/cm^3$.
The inset shows that the irreversible magnetization over a fuller temperature range.  In the inset, the x-symbols 
show the $30~K$ data and show that about $200~emu/cm^3$ or about $0.2~T$ flux is 
trapped at zero field.  At $6~K$, shown by the solid squares of the inset, 
the trapped flux at zero applied field for this rather porous 
granular pellet is $500~emu/cm^3$ or over $0.5~T$.

Using the Bean\cite {5} model with $J_c=17\Delta M/r$, the critical current densities of Fig. 5 
can be derived.  Here, $J_c$ is in $A/cm^2$, $\Delta M$  in $emu/cm^3$, and 
the sample radius, $r$, in $cm$.    
These $J_c$ values are not as high as a sintered pellet of 
$Nb_3Sn$,\cite {5} but the grain-to-grain supercurrent coupling is rather promising.  A scanning 
electron microscope picture of a fracture surface of the pellet shows grain size running from 
about $0.5~\mu m$ to $5~\mu m$.  The grains are somewhat faceted plates and many 
small grains are equiaxed.  From the magnitude of the measured screening currents, it is clear that 
the proper $r$ to put in the Bean model is the full sample diameter and not the grain size.  
The sample radius is about $1~mm$.  

Close to $T_c$, there is a reasonable range of thermodynamic reversibility, so an estimate 
has been made of $H_c$ vs. $T$.  To do this, it is assumed that the flux pinning in decreasing 
magnetic field is the same as in  increasing magnetic field and an equilibrium magnetization is 
defined as the average of the average of the increasing field magnetization and the decreasing field 
magnetization by $M_{eq}=(M_{inc}+M_{dec})/2$.  Willemin and co-workers\cite {6} have shown 
the validity of this procedure.  A plot of $M_{eq}~vs.~T$ in Fig.6 
shows behavior close to that predicted by Abrikosov.\cite {7} Integrating the area under these 
magnetization curves gives the $H_c~vs.~T$ curve shown in the inset of Fig. 6.  From the slope 
of $dH_c/dT=119~Oe/K$, one can calculate the jump in specific heat at $T_c$ to be 
$\Delta C=[VT_c/4\pi ][dH_c/dT]^2=79~mJ/mole~K$.  This is consistent with the direct 
specific heat measurements published earlier\cite {2} from which $\Delta C$ can be estimated to 
be about $84~mJ/mole~K$.

\section{Conclusions}

The transport and magnetization studies of $Mg^{10}B_2$ give very consistent measures of $H_{c2}(T)$ for this 
material with magnetization data giving 
a slope at $T_c$ of $dH_{c2}/dT=0.44~T/K$.  The magnetization 
$H_{c2}$ agrees with the onset of the resistive transition, and the broadened resistive transitions 
seem to reflect a flux-flow resistivity phenomenon.  Close to $T_c$, the values of the $H_c$ can be estimated and the slope is found to be $dH_c/dT=0.012T/K$.  This then means that the Ginsburg-Landau 
parameter, $\kappa =H_{c2}/[.707H_c]\simeq 26$.  By estimating 
$H_{c2}(T=0) = 0.71T_c[dH_{c2}/dT]_{T_c}$ to be $12.5~T$, the low temperature 
coherence distance, $\xi _o=[\phi _o/[2\pi H_{c2}]^{1/2}$ 
is found to be $5.2~nm$.  Using the 
relation $\kappa =\xi _o/\lambda $, the penetration depth, $\lambda =140~nm$.

The critical current densities for this rather porous sintered sample are on the order of  
$10^5~A/cm^2$ at $6~K$.  This would seem to indicate that $MgB_2$ grain boundaries 
can transmit rather large supercurrents.

\section{Acknowledgments}

We would like to thank M. J. Kramer and F. Laabs for the electron microscope pictures, 
and V. G. Kogan for useful discussions.  
C. Petrovic and 
C. E. Cunningham also made valuable contributions.
Ames Laboratory is operated for the
U. S. Department of Energy by Iowa State University under contract No.
W-7405-ENG-82 and
supported by the DOE, the Office of Basic Energy Sciences.  

*On leave from Commissariat a l'Energie Atomique, DRFMC-SPSMS, 38054 Grenoble, 
France.

\vfil\eject
\begin{figure}
\epsfxsize=0.9\hsize
\vbox{
\centerline{
\epsffile{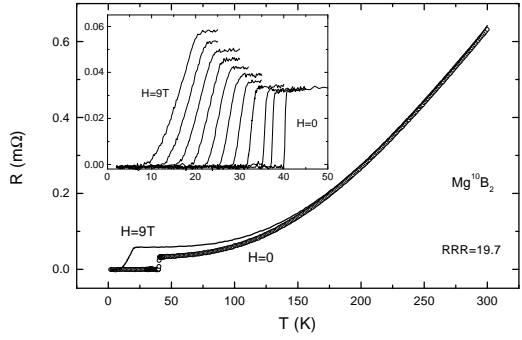}
}}
\caption{Resistance of $Mg^{10}B_2$ sample in zero and in $9~T$ applied field.  Inset: resistive 
superconducting transition in different applied fields (right to left): $0~T$, $0.5~T$, and 
from $1~T$ to $9~T$ in steps of $1~T$.} 
\label{F1}
\end{figure}
\begin{figure}
\epsfxsize=0.9\hsize
\vbox{
\centerline{
\epsffile{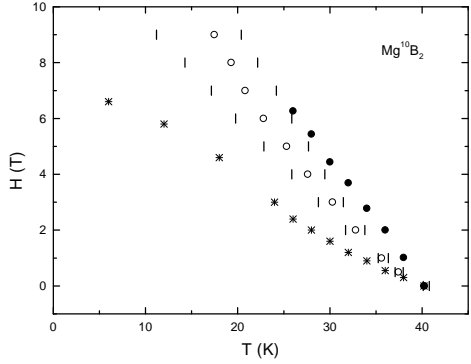}
}}
\caption{Upper critical field of $Mg^{10}B_2$ determined resistively (onset and offset are 
vertical bars and maximum slope points are the open circles) and from magnetization (filled 
circles).  Asterisks show $H_{irr}$.  }
\label{F2}
\end{figure}
\begin{figure}
\epsfxsize=0.9\hsize
\vbox{
\centerline{
\epsffile{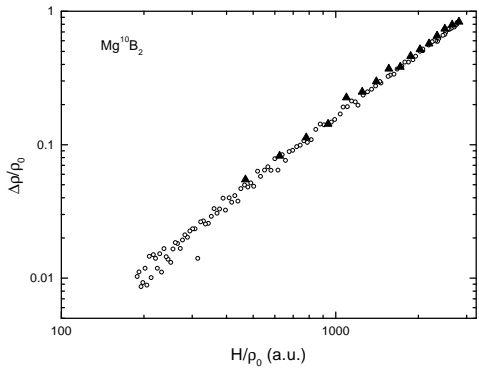}
}}
\caption{Kohler plot for $Mg^{10}B_2$: open circles from temperature-dependent 
resistance from $0~T$ to $9~T$ and filled triangles from field-dependent resistance at $45~K$. }
\label{F3}
\end{figure}
\begin{figure} 
\epsfxsize=0.9\hsize
\vbox{
\centerline{
\epsffile{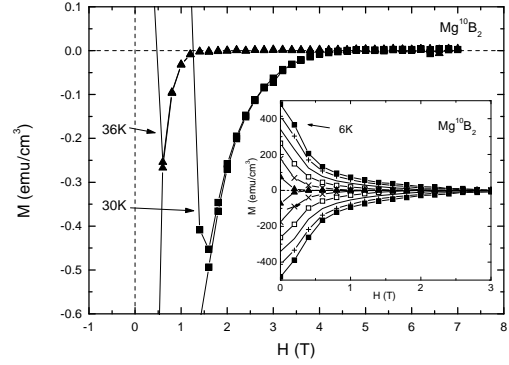}
}}
\caption{Expanded view of magnetization vs. field to show the reversible range and 
the $H_{irr}$ region.  
Inset shows the full range of magnetization up to $4~T$.}
\label{F4}
\end{figure}
\begin{figure}
\epsfxsize=0.9\hsize
\vbox{
\centerline{
\epsffile{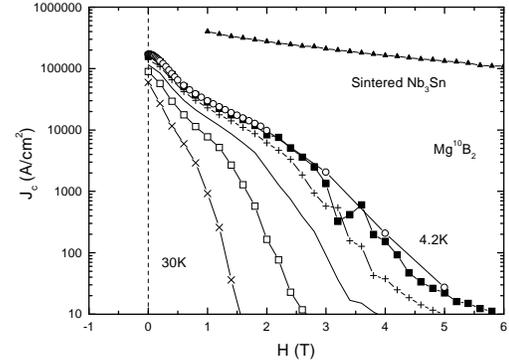}
}}
\caption{Comparison of $J_c(H,T)$  up to $30~K$ with published data for a sintered 
$Nb_3Sn$ sample at $4.2~K$.}
\label{F5}
\end{figure}
\begin{figure}
\epsfxsize=0.9\hsize
\vbox{
\centerline{
\epsffile{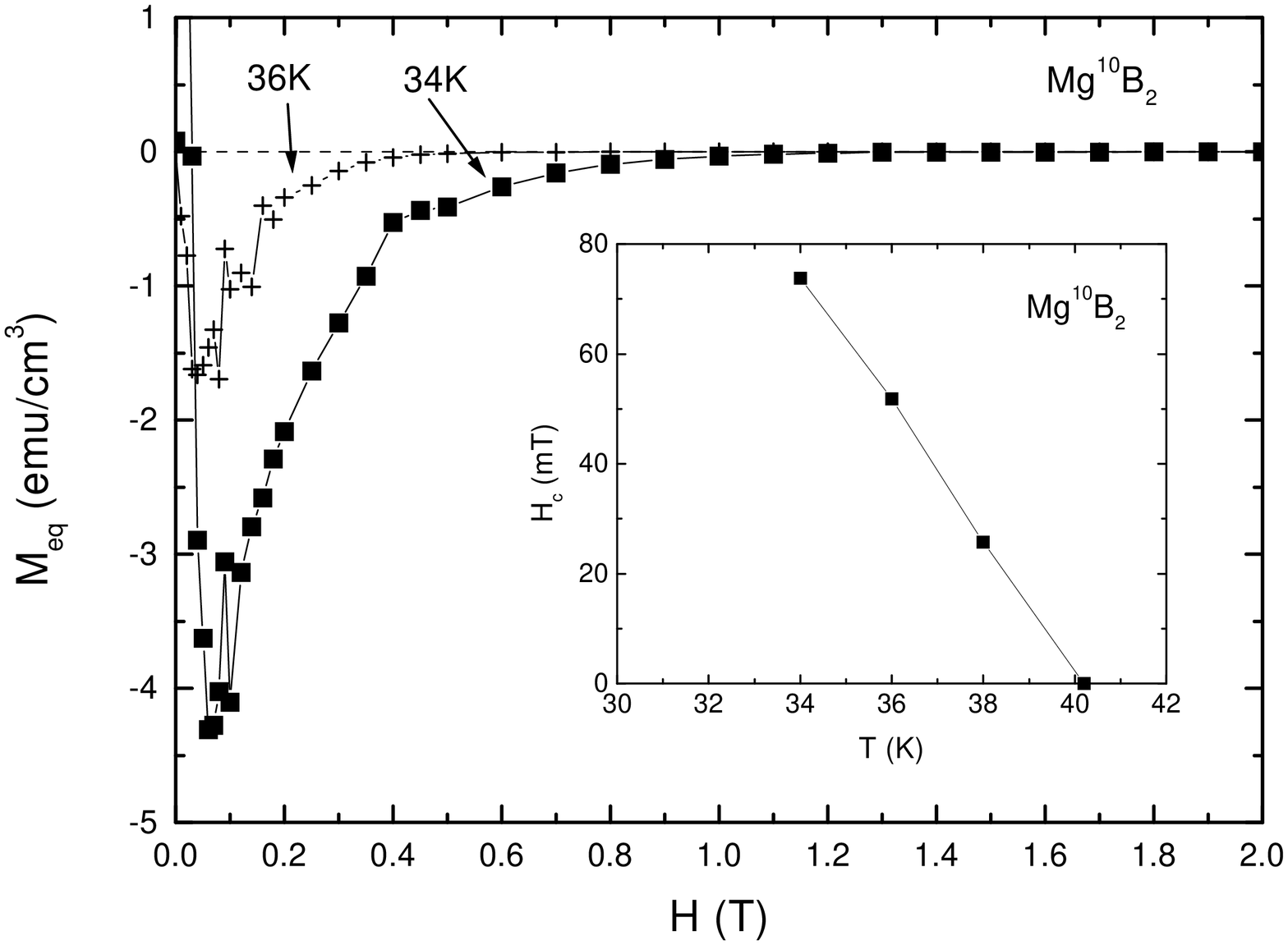}
}}
\caption{$M_{eq}~=~(M_{inc}+M_{dec})/2~vs.~T$ plot to show the extension of the 
reversible magnetization into the irreversible range assuming that flux pinning is the same for 
increasing and decreasing fields. The inset shows the resulting values of $H_c$. }
\label{F6}
\end{figure}

\vfil\eject

\end{document}